\documentclass[aip,jmp,preprint,prd,showpacs,nofootinbib]{revtex4-1}

\usepackage{latexsym}
\usepackage{amsmath}
\usepackage{graphicx}
\usepackage{subfigure}
\usepackage{dcolumn}
\usepackage{bm}
\usepackage{amssymb}
\usepackage{latexsym}
\usepackage{ulem}
\usepackage{color}
\usepackage[colorlinks,linkcolor=magenta,anchorcolor=blue,citecolor=blue]{hyperref}

\def\be{\begin{equation}}
\def\ee{\end{equation}}
\def\ba{\begin{eqnarray}}
\def\ea{\end{eqnarray}}

\begin{document}

\title{Effective Superpotentials of Type II D-brane/F-theory on Compact Complete Intersection Calabi-Yau Threefolds}

\author{Hao Zou}

\author{Fu-Zhong Yang}

\thanks{Corresponding Author}
\email{fzyang@ucas.ac.cn}

\affiliation{School of Physics, University of Chinese Academy of Sciences\\ Yuquan Road 19A, Beijing 100049, China}

\begin{abstract}
     In this paper, we extend the GKZ-system method to the more general case: compact Complete Intersection Calabi-Yau manifolds (CICY). For several one-deformation modulus compact CICYs with D-branes,  the on-shell superpotentials in this paper from the extended GKZ-system method are exactly consistent with published results obtained from other methods. We further compute the off-shell superpotentials of these models. Then we obtain both the on-shell and off-shell superpotentials for several two-deformation moduli compact CICYs with D-branes by using the extended GKZ-system method. The discrete symmetrical groups, $\mathbb{Z}_2$, $\mathbb{Z}_3$ and $\mathbb{Z}_4$, of the holomorphic curves wrapped by D-branes play the important roles in computing the superpotentials, in some sense, they are the quantum symmetries of these models. Furthermore, through the mirror symmetry, the Ooguri-Vafa invariants are extracted from the A-model instanton expansion.
\end{abstract}
\maketitle

\section{Introduction}
\indent In $\mathcal{N}=(2,2)$ superconformal field theory, mirror symmetry is an isomorphism between the $(c,c)$ and the $(a,c)$ chiral rings of the $\mathcal{N}=(2,2)$ superconformal algebra. In superstrings, mirror symmetry gives an equivalence between A-model compactified on X and B-model on Y, which can be explained as T-duality\cite{Strominger:1996tp}, and it opens a way to deal with non-perturbative topological string theory. The B-model is characterized  by complex structure moduli in terms of classical geometry, while the mirror A-model is characterized by K\"{a}hler moduli in terms of the quantum geometry. It is through mirror symmetry that correlation functions and non-perturbative prepotentials can be obtained exactly, which are know as the generating functions of Gromov-Witten invariants. In geometry,  the mirror symmetry relates two distinct Calabi-Yau manifolds: X and Y, in a way of exchanging their Complex moduli and K\"{a}hler moduli spaces\cite{Hori:2000wb,Skarke:1998vi}; and the  coefficients of the instanton expansion of the generating function can be interpreted as numbers of rational curves on the Calabi-Yau threefold\cite{Candelas:1991wu, Candelas:1991vp}.\\
\indent With the inclusion of D-branes and fluxes into Calabi-Yau threefold compactification, the $\mathcal{N}=2$ supersymmetry breaks into $\mathcal{N}=1$ supersymmetry. In the low energy regime, the type II string theory with the D-branes reduces to an effective $\mathcal{N}=1$ supergravity theory in four dimensions, and, accordingly, the moduli space is parameterized by $\mathcal{N}=1$ special geometry. In addition, including D-branes would introduce open string sectors and so the open string moduli. There exists two kinds of branes, the A- and B-branes, which wrap respectively on special Lagrangian submanifolds and holomorphic submanifolds on Calabi-Yau manifolds. On the B-model side, geometrically speaking, when varying the complex structure of Calabi-Yau space, a generic holomorphic curve will not be holomorphic with the respect to the new complex structure, and becomes obstructed to the deformation of the bulk moduli. The requirement for the holomorphy gives rise to a relation between the closed and open string moduli. Physically speaking, it turns out that the obstruction generates a superpotential for the effective theory depending on the closed and open string moduli. This superpotential is also defined as the F-term of low-energy effective theory, and it determines the string vacuum structure. The expression of instanton expansion of superpotential on the A-model side encodes the number of BPS states, and matematically it corresponds to the Ooguri-Vafa invariants, which are related to the open Gromov-Witten invariants and can be interpreted as counting holomorphic disks\cite{Aganagic:2000gs}. The algebraic geometric treatments of the open Gromov-Witten theory of a smooth projective variety X depend on
the moduli stack $\overline{\cal{M}}_{g,h,n} (X, C,\beta)$ which parametrizes the stable maps from genus g curves with h boundaries and n marked points to (X, C) with image class of the curve $\beta \in H_{2}(X, C, Z)$ and with image class of boundaries of the curve being in $ H_{1}(C, Z)$, where the submanifold C of X is wrapped by the D-branes from the physical viewpoint\cite{FLT:2012}. When $ h =0$, the moduli stack $\overline{\cal{M}}_{g,h,n} (X, C, \beta) $ just is the Kontsevich's moduli stack $\overline{\cal{M}}_{g,n} (X, \beta)$ which is a proper Deligne-Mumford stack with projective coarse moduli space\cite{BM:1996, KM:1994}. In particular, if X is a point, then $\overline{\cal{M}}_{g,n} (*)$ just is
the Deligne-Mumford moduli stack $\overline{\cal{M}}_{g,n}$ of stable curves\cite{Deligne Mumford:1969,Vistoli:1989}.

Similar to the closed string sector, open-closed mirror symmetry\cite{Vafa:1998tc} provides a powerful tool to compute the $\mathcal{N}=1$ superpotential. For the case of non-compact Calabi-Yau manifold, the D-brane effective superpotential can be computed by localizaton\cite{Aganagic:2000gs, Aganagic:2001nx, Aganagic:2003db}, topological vertex and direct integration related to $\mathcal{N}=1$ special geometry \cite{Lerche:2002ck,Lerche:2002yw}. For the compact cases, it was very complicate and difficult to apply these methods, especially for the compact Complete Intersection Calabi-Yau with multi-deformation moduli and including D-branes. To overcome the difficult, some techniques evolved, like mixed Hodge structure variation, Gauss-Manin connection\cite{Jockers:2009bf, Jockers:2008ef} and the blow-up method\cite{Grimm:2008ki,Grimm:2009ft,Grimm:2009ef,Grimm:2010cr}. Some work\cite{Alim:2009el,Alim:2010et,Xu:2013vi, Xu:2014tx,Cheng:2013hma,Zhang:2015uaa} have generalized the GKZ method of closed sectors\cite{Hosono:1993ik,Hosono:1994ax,Hosono:1995dd} to open-closed sectors, which gave a more effective way to calculate D-brane superpotentials. So far, these work mainly focused on the cases of Calabi-Yau hypersurfaces. The purpose of this paper is to further generalize this GKZ method to the more general case: the complete intersection Calabi-Yau manifold (CICY).\\
\indent In Toric geometry, compact Calabi-Yau threefolds are built as hypersurfaces or complete intersections defined by polyhedrons in ambient toric varieties. For open sectors, it is more comprehensive to consider the enhanced polyhedrons with one dimension higher, which can describe the configuration Calabi-Yau threefold and D-branes\cite{Li:2009wt, Alim:2009el}. This closely relates to the duality between type IIB theory compactified on Calabi-Yau threefolds and F-Theory on Calabi-Yau fourfolds which are elliptic fibration over Calabi-Yau threefolds. In the weak coupling limit $g_s \rightarrow 0$, the open-closed duality identifies the superpotential $\mathcal{W}$ of Type IIB theory compactified on Calabi-Yau threefold with the leading term of GVW flux superpotential $\mathcal{W}_{GVW}$ of F-theory on corresponding fourfold\cite{Berglund:2005dm,Jockers:2010um,Alim:2010et}:
	\begin{equation*}
		\mathcal{W}_{GVW}=\mathcal{W}+\mathcal{O}(g_s)+\mathcal{O}(e^{-1/g_s}).
	\end{equation*}
\indent The organization of this paper is as follows. In the next section, we review some aspects of superpotentials, relative periods, domain wall and Mirror map that will be used in this paper. In the third section, we first give a brief review of the Toric description of the Calabi-Yau manifold and generalized GKZ system, and then we further generalized it to describe compact Complete Intersection Calabi-Yau manifolds including D-branes. In the fourth section, applying the generalized GKZ method to some one-deformation modulus and  two-deformation moduli
 compact CICYs with D-branes, respectively, we obtain both on-shell and off-shell superpotentials and their Ooguri-Vafa invariants.
The discrete symmetrical groups, $\mathbb{Z}_2$, $\mathbb{Z}_3$ and $\mathbb{Z}_4$, of the holomorphic curves wrapped by D-branes play the important roles in computing the superpotentials, in some sense, they are the quantum symmetries of these models. The last section is a brief summary.\\

\section{Superpotential and Relative Period}
\subsection{Effective Superpotential}
\indent The effective superpotential consists of two parts: the D-brane superpotental and the background flux superpotential:
	\be
		\mathcal{W}(z,\hat{z})=\mathcal{W}_{brane}(z,\hat{z})+\mathcal{W}_{flux}(z).
	\ee
\indent On the B-model side, the effective superpotential arises from D5-brane warping 2-cycle $C$ on Calabi-Yau threefold, Y, is
	\be
		\mathcal{W}_{brane}=N_{\hat{a}}\int_{\hat{\Gamma}^{\hat{a}}(\hat{z})} \Omega^{3,0}(z,\hat{z})=\sum N_{\hat{a}}{\hat{\Pi}}^{\hat{a}}(z,\hat{z}),
	\ee
where $\hat{\Gamma}^{\hat{a}} \in H_3(Y,C)$ are basis of three chains with non-trivial boundaries $\partial \hat{\Gamma}^{\hat{a}} \supset C$, and ${\hat{\Pi}}^{\hat{a}}(z,\hat{z})$ are called semi periods, which are the integrals of the holomorphic $(3,0)$-form $\Omega^{3,0}(z,\hat{z})$ over $\hat{\Gamma}^{\hat{a}}$\cite{Lerche:2002ck, Jockers:2008ef}.\\
\indent The internal background fluxes $F=F_{RR}+\tau F_{NS}$ leads to an $\mathcal{N}=1$ superpotential in closed-string sector, and it also could be written in terms of periods $\Pi^a(z) \in H_3(Y)$:
	\be
		\mathcal{W}_{flux}(z)=\int_Y F \wedge \Omega^{3,0}(z,\hat{z})=\sum N_a \Pi^a(z).
	\ee
\indent Hence the combined effective superpotential written in terms of linear combination of relative period is:
	\be
		\mathcal{W}(z,\hat{z})=\sum \underline{N}_{\alpha}\underline{\Pi}^{\alpha}(z,\hat{z})
	\ee
where the coefficients $\underline{N}_{\alpha}$ are determined by both the topological data of the brane and flux background, and $\underline{\Pi}^{\alpha}(z,\hat{z})$ are unified expressions capturing both the semi-period integrals of $\Omega^{3,0}$ over 3-chain $\hat{\Gamma}^{\hat{z}}$ whose boundaries are wrapped by D-branes and the period integrals over 3-cycles. These relative periods can be written as integrals of relative forms $\underline{\Omega}(z,\hat{z})$ over relative homology basis $\underline{\Gamma}^{\alpha}\in H_3(Y,C)$:
	\be
	\label{rel_per}
		\underline{\Pi}^{\alpha}(z,\hat{z})=\int_{\underline{\Gamma}^\alpha(z,\hat{z})}\underline{\Omega}(z,\hat{z}).
	\ee
\indent The above superpotential is called off-shell superpotential, since it depends on both closed- and open-moduli. The open-moduli space is the parameter space of deformations of the 2-cycle. In most situations, we're more care about the on-shell case along the open-string condition, {\it i.e.} fixing the open parameter. Then the on-shell condition is
	\be
	\label{osc}
		\dfrac{d}{d\hat{z}}\mathcal{W}(z,\hat{z})=0,
	\ee
which leads to the on-shell superpotential $\left.W(z)=\mathcal{W}(z,\hat{z})\right|_{\hat{z}=\rm {critical\ value}}$.\\
\indent One thing needed to point out is that, when the 2-cycle $C$ is holomorphic, it will give the BPS states of D-branes.  However,  when $C$ is non-holomorphic after certain deformations, and then it will correspond to non-BPS states. The trick to deal with this is to embed the 2-cycle $C$ in a holomorphic ample divisor $D$ so as to consider a much simpler homology group $H_3(Y,D)$ of an unobstructed theory. For more details, we recommend to refer \cite{Lerche:2002yw, Lerche:2002ck, Jockers:2008ef,Alim:2010et}\\

\subsection{Relative Period and Domain Wall}
In Eq.(\ref{rel_per}), it's usually to take the boundary of $\underline{\Gamma}^\alpha(z,\hat{z}) \in H_3(Y,D)$ ending on two 2-cycle:
	\be
	\partial\underline{\Gamma}(z,\hat{z})=C^+(z,\hat{z})-C^-(z,\hat{z}),\  [C^+]=[C^-]\in H_2(Y).
	\ee
The domain wall tension, denoted as $\mathcal{T}(z,\hat{z})$, is the difference between the values of off-shell superpotentials for two D-brane wrapping $C^+$ and $C^-$, respectively:
	\be
	\label{dwt}
		\mathcal{T}(z,\hat{z})=\mathcal{W}(C^+)-\mathcal{W}(C^-).
	\ee
In this off-shell case, it equals to the relative period: $\mathcal{T}(z,\hat{z})=\underline{\Pi}(z,\hat{z})$ which could be explained by the Abel-Jacobi map\cite{Aganagic:2000gs, Li:2009wt}. When $\hat{z}$ take critical values,  $T(z)=\mathcal{T}(z,\hat{z}_{\rm critic})$ is differed from $\underline{\Pi}(z,\hat{z}_{\rm critic})$ by bulk periods, but Eq.(\ref{dwt}) still holds.\\
\indent Than there is an alternative definition of on-shell condition({\it i.e. }Eq.(\ref{osc})) in the open-string direction by Noether-Lefshetz locus in mathematics\cite{Clemens:1998wm,Alim:2010et}:
	\be
	\label{osc2}
		\mathcal{N}=\left\{(z,\hat{z})\left|\dfrac{d\underline{\Pi}(z,\hat{z})}{d\hat{z}}=0\right.\right\}=\left\{(z,\hat{z})\left|\pi(z,\hat{z};\partial\underline{\Gamma}(z,\hat{z}))=0\right.\right\},
	\ee
where $\pi$ is the period vector of the ample divisor $D$. This definition provides a method of integrating periods on a reduced subsystem of one-dimension lower to obtain domain wall tension.
\subsection{Picard-Fuchs Equation}
\indent Analogy to the closed-string sector, the relative periods also satisfy a set of differential equations named Picad-Fuchs equations. The Picard-Fuchs equations have the following form:
	\be
	\label{pf_eq1}
		\mathcal{L}_a(\theta,\hat{\theta})\underline{\Pi}(z,\hat{z})=0,
	\ee
where $\theta$ and $\hat{\theta}$ are abbreviations of $z\partial_z$ and $\hat{z}\partial_{\hat{z}}$. The explicit form of these operators could be derived from Gauss-Manni system, which is the result of variation of mixed Hodge structure. The solutions of this differential system lie in the generalized hypergeometric GKZ system, which would be discussed in the next section. The differential operators have two parts:
	\be
		\mathcal{L}_a(\theta,\hat{\theta})=\mathcal{L}^b_a(z)-\mathcal{L}^{bd}_a(z,\hat{z})\hat{\theta},
	\ee
where the superscript $b$ stands for bulk, $bd$ stands for boundray, and both $\mathcal{L}^b_a(z)$ and $\mathcal{L}^{bd}_a(z,\hat{z})$ vanish bulk periods. And in the relative form level, the r.h.s. of Eq.(\ref{pf_eq1}) differs from zero by at most a differential of a two-form $\omega^{2,0}$: $\mathcal{L}_a(\theta,\hat{\theta})\underline{\Omega}=d\omega$. From last subsection, we have:
	\be
	\label{pf_eq2}
		\mathcal{L}_a(\theta,\hat{\theta})\mathcal{T}(z,\hat{z})=0,\ i.e.\ \mathcal{L}^b_a\mathcal{T}(z,\hat{z})=\mathcal{L}^{bd}_a\mathcal{T}(z,\hat{z}).
	\ee
\indent Combined with Eq.(\ref{osc2}), the periods on subsystem from ample divisor $D$ satisfy
	\be
	\label{subper}
		2 \pi i \hat{\theta} \mathcal{T}(z,\hat{z}) = \pi(z,\hat{z}),
	\ee
which in turn offer a way to calculate domain wall tension by integrating. Then substitute it into Eq.(\ref{pf_eq2}) and fix $\hat{z}$ at critical value, we obtain the inhomogeneous Picard-Fuchs equations only with bulk differential operators:
	\be
	\label{ihpf}
		\mathcal{L}^b_a T(z) =f_a(z),
	\ee
where $f_a(z)$ are defined as $2 \pi i f_a(z)=\mathcal{L}^{bd}_a\pi(z,\hat{z}_{\rm critic})$. When applying it to concrete example, one can find the inhomogeneous term of the r.h.s of above equation are coincide with results in \cite{Walcher:2009di, Alim:2010et}. This offers a method to check such system.
\subsection{Mirror Map and Instanton Sum}
\indent From above one can obtain both off-shell and on-shell superpotential in B-model and relative periods, $\underline{\Pi}^{\alpha}$, which are solutions of Picard-Fuchs equations. Through mirror map, one can calculate corresponding instanton expansion in A-model, {\it i.e.} the flat coordinates in A-model at large radius regime are related to the flat coordinates of B-model at complex structure regime. Then the closed- and open-flat coordinates on the A-model side are defined as:
	\be
		t^a(z) = \frac{\underline{\Pi}^a(z)}{\underline{\Pi}^0(z)},\quad \hat{t}^{\hat{a}}=\frac{\underline{\Pi}^{\hat{a}}(z,\hat{z})}{\underline{\Pi}^0(z)},
	\ee
where $\underline{\Pi}^0(z)$, $\underline{\Pi}^a(z)$ and $\underline{\Pi}^{\hat{a}}(z,\hat{z})$ are chosen basis of relative periods associated with closed- and open-moduli. The relative period vector in flat coordinates could be written as
	\be
		\underline{\Pi}^\alpha(t,\hat{t})=\Pi^0\left(1,\ t^a,\ \partial_{t^a}\mathcal{F}(t),\ 2\mathcal{F}(t)-\sum t^a\partial_{t^a}\mathcal{F}(t);\ \hat{t}^{\hat{a}},\ \mathcal{W}^{\pm}(t,\hat{t}),\ \ldots \right).
	\ee
In this expression, $\mathcal{F}(t)$ is the closed-string holomorphic $\mathcal{N}=2$ prepotential and $\mathcal{W}^{\pm}$ correspond to superpotential. The instanton corrections are encoded as a power series expansion of $q^a = e^{t^a}$ and $\hat{q}^{\hat{a}}=e^{\hat{t}^{\hat{a}}}$:
	\be
	\label{ins}
		\mathcal{W}(t,\hat{t})=\sum_{\vec{k},\vec{m}}G_{\vec{k},\vec{m}}q^{\vec{k}}\hat{q}^{\hat{m}}=\sum_n \sum_{\vec{k},\vec{m}}\frac{N_{\vec{k},\vec{m}}}{n^2}q^{n\vec{k}}\hat{q}^{n\vec{m}}.
	\ee
\indent In Eq.(\ref{ins}), $\left\{G_{\vec{m},\vec{k}}\right\}$ are open Gromov-Witten invariants labeled by relative homology class, where $\vec{m}$ represent the elements of $H_1(D)$ and $\vec{k}$ represent the elements of $H_2(Y)$, and $\left\{N_{\vec{k},\vec{m}}\right\}$ are Ooguri-Vafa invariants.
\section{CICY in Toric Geometry and Extended GKZ System}
\indent In general, the complete intersection of $l$ hypersurfaces determined by $l$ polynomial $p_r$ ($r=1,\ldots,l$) in the product of $k$ weighted projective spaces is denoted by:
	\be
	\left(
		\begin{array}{c}
		\mathbb{P}^{n_1}[\omega_1^{(1)},\cdots,\omega_{n_1+1}^{(1)}]\\
		\vdots\\
		\mathbb{P}^{n_k}[\omega_1^{(k)},\cdots,\omega_{n_k+1}^{(k)}]\\
		\end{array}
	\right|
	\left|
		\begin{array}{ccc}
		d_1^{(1)}&,\cdots,&d_l^{(1)}\\
		\vdots&\  &\vdots\\
		d_1^{(k)}&,\cdots,&d_l^{(k)}\\
		\end{array}
	\right)_{\chi}^{h^{1,1}},
	\ee
where $d^{(i)}_r$ is the degree of the coordinates of $\mathbb{P}^{n_i}[\omega^{(i)}_1,\ldots,\omega^{(i)}_{n_i+1}]$ in the $r$-th polynomial $p_r$ ($i=1,\ldots,k$;\ r=1,\ldots,l). In order to be a Calabi-Yau manifold, $\omega^{(i)}_j$ and $d^{(i)}_r$ should satisfy the Calabi-Yau condition:
	\be
	\label{cy_cond}
		\sum_{j=1}^{n_i+1}\omega^{(i)}_j-\sum_{r=1}^{l}d^{(i)}_r \equiv 0.
	\ee
\indent For convenience, we only consider complete intersection examples in non-singular ambient spaces in this paper.
Since the ambient space is a product space $\mathbb{P}^{n_1}[\vec{\omega}^{(1)}]\times \dots \times\mathbb{P}^{n_k}[\vec{\omega}^{(k)}]$, it can be described by the reflexive polyhedra $\Delta$ in $\mathbb{R}^{n_1} \times \ \dots \times \mathbb{R}^{n_k}$. Then the CICY manifold is constructed from this reflexive polyhedra $\Delta$ by nef-partition, while the mirror CICY manifold is determined by dual reflexive polyhedra $\Delta^*$ with dual nef-partition\cite{Borisov:1993tc,Batyrev:1994va,Batyrev:1994wc,Klemm:2004dl}.\\
\indent A nef-partition is a decomposition of the vertex set $V$ of $\Delta^{*}$ into a disjoint union
	\be
	V=V_1 \sqcup V_2 \sqcup \dots \sqcup V_{l}
	\ee
such that divisors $E_i =\sum\limits_{v \in V_i} D_v$ are Cartier,where $D_v$ are prime torus-invariant Weil divisors corresponding to vertices of $\Delta^{*}$. Let $\Delta_i $ be the convex hull of vertices from $V_i \cup \{0\}$. Then these polytopes form a nef-partition if their Minkowski sum $\Delta$ is a reflexive polytope. The dual nef-partition is formed by polytopes $\Delta_i$ of $E_i$, which give a decomposition of the vertex set of $\nabla^{*}$ and their Minkowski sum is $\Delta$. One can find that the duality of reflexive polytopes just switches convex hull and Minkowski sum for dual nef-partitions:
	\be
	\begin{array}{ccc}
		\Delta = \Delta_1 + \Delta_2+ \dots +\Delta_{l} & \overset{dual}{\longleftrightarrow} &\Delta^{*} = {\rm Conv}(\nabla_1,\ \nabla_2,\ldots, \nabla_{l})\\
		\nabla^{*} = {\rm Conv}(\Delta_1,\ \Delta_2,\ldots,\ \Delta_{l}) & \overset{dual}{\longleftrightarrow}  & \nabla = \nabla_1+\nabla_2+\dots+\nabla_{l}\\
	\end{array}
	\ee
\indent In our examples, the vertices of $\Delta^*$ are the integral points $v^{*}_{i,1}=(1,0,\dots,0),\ldots,v^{*}_{i,n_i}=(0,\dots,0,1)$ and $v^{*}_{i,n_i+1}=(-\omega^{(i)}_1,\dots,-\omega^{(i)}_{n_i})$. These vertices are decomposed into $l$ sets $V_r$ (r=1,\ldots,l) and each $V_r$ contains $d^{(i)}_r$ vertices from $\left\{v^*_{i,j}\right\}_{1\leq j \leq n_i+1}$ for each $i=1,\ldots,k$. Then $l$ defining polynomials can be written in toric coordinates $X_{m,n}$($m=1,\ldots,k;\ n=1,\ldots,n_m$):
	\be
	\label{intersections}
		P_r = \sum_{v_{i,j} \in V_r \cup \{0\}} a_{i,j}X^{v^*_{i,j}},\quad r=1,\dots,l
	\ee
\indent It is convenient to extend each vertices to $\left\{\overline{v}^*_{i,j}=(\vec{e}^{(r)},v^*_{i,j})\right\}$ in $\mathbb{R}^l\times\mathbb{R}^{n_1}\dots\times\mathbb{R}^{n_k}$, where $\vec{e}^{(r)}$ is the $r$-th unit vector of $\mathbb{R}^l$, and introduce $l$ extra vertices $\left\{\overline{v}^*_{0,r}=(\vec{e}^{(r)},\vec{0})\right\}$. Then one can determine $k$ generators named $l^{(s)}$-vectors of relation lattice
	\be
		\mathbb{L}=\left\{\langle l^{(s)}\rangle_{s=1,\dots,k}\left| \sum\limits_{\rm all\ vertices} l^{(s)} \overline{v}^*_{i,j}=0\right. \right\},
	\ee
 where $l^{(s)}$ in proper chosen basis can be written as:
	\be
		l^{(s)}=\left(-d^{(s)}_1,\ldots,-d^{(s)}_l;0,\ldots,0,\omega^{(s)}_1,\ldots,\omega^{(s)}_{n_s+1},0,\ldots\right)\equiv \left(\left\{l^{(s)}_{0_i}\right\};\left\{l^{(s)}_j\right\}\right),
	\ee
where they automatically satisfy the Calabi-Yau condition(\ref{cy_cond}). One thing need to note is that this space is always with singularities and should be resolved by blowing up: adding new vertices in the polytope being $\hat{\Delta}$. But the relation lattice is still the same and so are the generators. Then the moduli variable associated to $l^{(s)}$ is given by
	\be
		z_s =(-1)^{\sum l^{(s)}_{0_i}}a^{l^{(s)}}=(-1)^{\sum_{i=1}^ld^{(s)}_i}\dfrac{a_{s,1}^{\omega^{(s)}_1}\dots a_{s,n_s+1}^{\omega^{(s)}_{n_s+1}}}{a_{1,0}^{d^{(s)}_1} \dots a_{l,0}^{d^{(s)}_l}}.
	\ee

\indent The above is the story about closed-string sector on CICY, one can find more details about corresponding closed GKZ system, Picard-Fuchs equations as well as their solutions in \cite{Hosono:1995dd,Hosono:1993ik,Hosono:1994ax,Klemm:2004dl}. On the other hand, the open-string sector from D-branes can be described by the family of hypersurfaces $D$, which is defined as intersections $\{P=0\}\cap\{Q(D)=0\}$. In toric language, the $Q(D)$ is defined by $p$ vertices $\overline{v}^*_{i,j}=(\vec{e}^{(r)},v^*_{i,j})$ chosen from $V\cup\{0\}$ (Here we prefer to replace notations by $\overline{w}^*_{i}=(\vec{e}^{(r)},w^*_{i,j})\ (i=1,\ldots,p)$ for disambiguation):
	\be
	\label{divisor}
		Q(D)=\sum_{i=1}^{p}b_{i}X^{w^*_i}.
	\ee
In this paper, we mainly consider about one open-string moduli $\hat{z}$, and the vertices determining $Q(D)$ comes from the same nef-partition $V_i\cup \{0\}\ (i=1,\ldots,l)$. Then Eq.(\ref{divisor}) has the form $Q(D)=X^{v^*_{i,j}}+\hat{z}X^{{v^*_{i,j}}^{\prime}}$.\\
\indent It turns out that we can also consider enhanced polyhedron with one-dimension higher, which could give a comprehensive description for both the ambient CICY and specific chosen divisor. It also corresponds to dual F-theory compactified on fourfold determined by this enhanced polyhedron. Then the mirror manifold together with brane geometry can be constructed by dual reflexive enhanced polyhedrons $(\underline{\Delta},\underline{\Delta}^*)$ as $\{\underline{v}_{i,j}^*\}$ and $\{{\underline{w}}_{i,j}^*\}$ in product space $\mathbb{R}^l\times\mathbb{R}^{n_1}\times\cdots\times\mathbb{R}^{n_k}\times\mathbb{R}^*$, where $\{\underline{v}_{i,j}^*\}=\{(\overline{v}^*_{i,j};0)\}=\{(\vec{e}^{(m)};v_{i,j}^*;0)\}$ determine the $l$ hypersurfaces function, and $\{\underline{w}_{i}^*\}=\{(\overline{w}^*_i;1)\}$ describe the divisor we would choose to calculate. \\
\indent Accordingly, the generator of the extended relation lattice $\underline{\mathbb{L}}$ can be write in three parts:
	\be
		\underline{l}^{(s)}=\left(\left\{\underline{l}^{(s)}_{0_i}\right\};\left\{\underline{l}^{(s)}_{c,j}\right\};\left\{\underline{l}^{(s)}_{o,j}\right\}\right)
	\ee
where the first part $\{\underline{l}^{(s)}_{0_i}\}$ is related to nef-partition, and note that if  $\{\underline{l}^{(s)}_{0_i}\}$ has only one component valued $1$, i.e. $i$ has only one value, it degenerates to the hypersurface case in which we usually omit this part; the second part $\{\underline{l}^{(s)}_{c,j}\}$ and the third part $\{\underline{l}^{(s)}_{o,j}\}$ correspond to closed- and open-sector respectively. The moduli variables are given by
	\be
		z_s=(-1)^{\sum l^{(s)}_{0_i}}a^{\underline{l}^{(s)}}=(-1)^{\sum l^{(s)}_{0_i}}\prod_{i,j} a_{i,j}^{\underline{l}^{(s)}_{i,j}}.
	\ee
Here we used the same notation as before may cause some ambiguity, but it's easy to distinguish in concrete example, and using two variables, $\{i,j\}$, to parameter $l^{(s)}$'s components is convenient to correspond $\{a_{i,j}\}$ from $l$ equations(\ref{intersections}). Then differential operators for Picard-Fuchs equations $\mathcal{L}^{(s)}\underline{\Pi}(z,\hat{z})=0$ from this extended GKZ system have following form:
	\be
	\label{gkz}
		\mathcal{L}^{(s)}=\prod_{i=1}^{l}\prod_{k=1}^{\underline{l}^{(s)}_{0_i}}(\theta_{0_i}-k)\prod_{\underline{l}^{(s)}_{i,j}>0}\prod_{k=0}^{\underline{l}^{(s)}_{i,j}-1}(\theta_{i,j}-k)-(-1)^{\sum l^{(s)}_{0_i}}z_s\prod_{i=1}^{l}\prod_{k=1}^{-\underline{l}^{(s)}_{0_i}}(\theta_{0_i}-k)\prod_{\underline{l}^{(s)}_{i,j}<0}\prod_{k=0}^{-\underline{l}^{(s)}_{i,j}-1}(\theta_{i,j}-k)
	\ee
where $\theta_{i,j}=a_{i,j}\dfrac{\partial}{\partial a_{i,j}}$ and is related to $\Theta_{z_s}=z_s \dfrac{\partial}{\partial z_s}$ by
	\be
		\theta_{i,j}=\sum_{s=1}^l \underline{l}^{(s)}_{i,j} \Theta_{z_s}
	\ee
\indent The solutions to this extended GKZ system(\ref{gkz}) can be written immediately as:
	\be
		B_{\{l^{(s)}\}}(\{z_s\};\{\rho_s\})=\sum_{\{n_s\}}\dfrac{\prod_i \Gamma(1-\sum_s l^{(s)}_{0_i} (n_s+\rho_s))}{\prod_j \Gamma(1+\sum_s l^{(s)}_{c,j}(n_s+\rho_s)) \prod_j \Gamma(1+\sum_s l^{(s)}_{o,j}(n_s+\rho_s))}\prod_s z_s^{n_s+\rho_s}
	\ee
\indent Take $\omega_0(\{z_s\})=\left.B_{\{l^{(s)}\}}(\{z_s\};\{\rho_s\})\right|_{\{\rho_s=0\}}$ and $\omega_i(\{z_s\})=\left.\partial_{\rho_i}B_{\{l^{(s)}\}}(\{z_s\};\{\rho_s\})\right|_{\{\rho_s=0\}}$, we could obtain the open-closed mirror map for this comprehensive system of type II theory as well as for the dual F-theory:
	\be
		t_i(\{z_s\})=\dfrac{\omega_i(\{z_s\})}{\omega_0(\{z_s\})}.
	\ee
Similarly to Eq.(\ref{ins}), the superpotential, as a chosen special solution to Picard-Fuchs equations, encodes the Gromov-Witten invariants as well as Ooguri-Vafa invariants in power series of $q_i=e^{t_i}$.
\section{ Superpotentials and Ooguri-Vafa Invariants}
In this subsection and next subsection, we will use the GKZ system method to calculate the on-shell and off- shell superpotentials and the Ooguri-Vafa invariants for some one-deformation modulus and two-deformation moduli compact CICYs with D-branes. The on-shell results of one-deformation modulus compact CICYs with D-branes are agree with those in the papers \cite{Walcher:2009di,Aganagic:2009jq} obtained from the direct integration method. We further compute the off-shell superpotentials of these models. Then we obtain both the on-shell and off-shell superpotentials for several two-deformation moduli compact CICYs with D-branes by using the extended GKZ-system method. The discrete symmetrical groups, $\mathbb{Z}_2$, $\mathbb{Z}_3$ and $\mathbb{Z}_4$, of the holomorphic curves wrapped by D-branes play the important roles in computing the superpotentials, in some sense, they are the quantum symmetries of these models. Furthermore, through the mirror symmetry, the Ooguri-Vafa invariants are extracted from the A-model instanton expansion.
\subsection{One-deformation Modulus  Compact CICYs with D-branes}
In this subsection, we focus on three examples with one parameter, with D-branes wrapped on curves characterized by $\mathbb{Z}_2$-, $\mathbb{Z}_3$- and $\mathbb{Z}_4$-symmetry respectively.\\
\\
\noindent{\it 1}. {\textbf{$\mathbb{P}^5[3,3]$}}\\
\indent This is the intersection of two cubics in $\mathbb{C}P^5$, denoted by
	\be
		Y=\{P_1=0, P_2=0\}/G.
	\ee
In Toric description, this intersection could be determined by following vertices after Nef-Partition:
	\be
		\begin{split}
			&E_1=\{ \nu^*_1=(1,0,0,0,0);\nu^*_2=(0,1,0,0,0);\nu^*_3=(0,0,1,0,0)\} \\
			&E_2=\{ \nu^*_4=(0,0,0,1,0);\nu^*_5=(0,1,0,0,1);\nu^*_6=(-1,-1,-1,-1,-1) \}\\
		\end{split}
	 \ee
The charge vector is given by
	\be
		l=(-3,-3;1,1,1,1,1,1).
	\ee		
And the invariant complex structure parameter, which is the closed string moduli here, is given by $z_1\equiv a^{l}=\dfrac{a_{1,1} a_{1,2} a_{1,3} a_{2,1} a_{2,2} a_{2,3}}{a_{1,0}^3 a_{2,0}^3}$. Then we have following two functions determining Y in Toric Coordinates:
	\be
		\left \{
		\begin{split}
			&W_1(a,X)=a_{1,0}+a_{1,1} X_1+a_{1,2} X_2+a_{1,3} X_3\\
			&W_2(a,X)=a_{2,0}+a_{2,1} X_4+a_{1,2} X_5+a_{2,3} (X_1 X_2 X_3 X_4 X_5)^{-1}\\
		\end{split}
		\right.
	\ee
After coordinates transformation as:
	\be
		X_1=\frac{x_1^3}{x_4 x_5 x_6} \quad
		X_2=\frac{x_2^3}{x_4 x_5 x_6} \quad
		X_3=\frac{x_3^3}{x_4 x_5 x_6} \quad
		X_4=\frac{x_4^3}{x_1 x_2 x_3}\quad
		X_5=\frac{x_5^3}{x_1 x_2 x_3}\quad
	\ee
We obtain cubic polynomials in homogeneous coordinates:
	\be
		\left \{
		\begin{split}
			&P_1=x_1^3+x_2^3+x_3^3+\psi x_4 x_5 x_6\\
			&P_2=x_4^3+x_5^3+x_6^3+\psi x_1 x_2 x_3\\
		\end{split}
		\right.
       	\ee
where $\psi=z_1^{-\frac{1}{6}}$. \\
\indent Following J. Walcher \cite{Walcher:2009di}, we also choose these two curves, denoted by $\mathcal{C}_\pm$, given by
	\be
		\{x_1=\eta x_2,\ x_4=\zeta x_5,\ x_3^3+\psi \zeta x_5^2x_6=0,\ x_6^3+\psi \eta x_2^2x_3=0\}
	\ee
where $\eta^2=-1,\ \zeta^2=-1$. \\
\indent To calculate the domain wall tensions and the superpotentials for the vacua  $C_\pm$, we could study the divisor: $Q(D)=x_1^3+z_2 x_2^3$, which interpolates between the two vacua. Here $z_2$ is the open string moduli, and according to this divisor the critical point of vacua is at $z_2=1$.\\
\indent In the corresponding enhanced polyhedron, the vertices are $E_1=\{\underline{\nu}^*_i=(1,0;{\nu}^*_i;0)\}_{i=1,2,3}$ and $E_2=\{\underline{\nu}^*_i=(0,1;{\nu}^*_i;0)\}_{i=4,5,6}$, and so the chosen divisor is determined by following two vertices:
	\be
		\underline{w}_{1}^*=(1,0; 1,0,0,0,0;1),\ \underline{w}_{2}^*=(1,0; 0,1,0,0,0;1)
	\ee
Intersect with space Y, namely $\{P_1=0, P_2=0\}\cap\{ Q(D)=0 \}$. By $x_2=(-z_2)^{-\frac{1}{3}}x_1$, we get the subsystem:
	\be
   		\left \{
		\begin{split}
			&P_1^D=x_1^3+x_3^3+\widetilde{\psi} x_4 x_5 x_6\\
			&P_2^D=x_4^3+x_5^3+x_6^3+\widetilde{\psi}  x_1^2 x_3\\
		\end{split}
		\right.
	\ee
where $\widetilde{\psi} =u^{-\frac{1}{6}},\ u=-(\dfrac{z_1}{z_2})(1-z_2)^2$. And the corresponding charge vector become
	\be
		\tilde{l}=(-3,-3;2,1,1,1,1).
	\ee	
From the GKZ system, we can compute the periods on this subsystem. First, the Picard-Fuchs equation is given by
	\begin{equation*}
		\mathcal{L}^D=2\theta (2\theta-1)\theta^4-u(3\theta-1)^2(3\theta-2)^2(3\theta-3)^2
	\end{equation*}
where $\theta=u\dfrac{\partial}{\partial u}$. Through Frobenius method, the regular solution with fractional power in $u$ is:
	\be
		\pi(u;\frac{1}{2})=\frac{c}{2}B_{\{\widetilde{l}\}}(u;\frac{1}{2})=\frac{c}{2}\sum_{n=0}^{\infty}\frac{\Gamma(1+3(n+\frac{1}{2}))^2u^{n+\frac{1}{2}}}{\Gamma(1+n+\frac{1}{2})^4\Gamma(1+2(n+\frac{1}{2}))}
	\ee
where $B_{\{l^a\}}(u_a;\rho_a)$ is the generating function of solutions of GKZ system, $\dfrac{c}{2}$ is the normalization constant\footnote{The $1/2$ in the normalization constant is chosen for latter computation's convenience.}. Then we could compute the domain wall tensions from $2\pi i\hat{\theta}\mathcal{T}(z,\hat z)=\pi(z,\hat z)$, which is dependent on $\xi:=\sqrt{z_2}$ on this subsystem, as
	\be
		\mathcal {T}(z_1,z_2)=\frac{1}{2\pi i}\int\pi (z_2) \frac{{\rm d}z_2}{z_2}=\frac{1}{2\pi i}\int\pi (\xi) \frac{2{\rm d}\xi}{\xi}
	\ee
Due to $\mathcal{T}=\mathcal{W}^+-\mathcal{W}^-$ and the $\mathbb{Z}_2$-Symmetry between $\mathcal{C}_+$ and $\mathcal{C}_-$, namely $\mathcal W^+=-\mathcal W^-$, we could compute following integration
	\be
	\begin{split}
		4\pi i\mathcal W^+&=\int_{-\sqrt{z_2}}^{+\sqrt{z_2}}\pi (\xi) \frac{{\rm d}\xi}{\xi}\\
		&=\frac{c}{2}\left.{\sum_{n=0}^{\infty}\frac{\Gamma(1+3(n+\frac{1}{2}))^2(-z_1)^{n+\frac{1}{2}}{}_2F_1(-1-2n,-\frac{1}{2}-n,\frac{1}{2}-n,\xi^2)}{(1+2n)\Gamma(1+n+\frac{1}{2})^4\Gamma(1+2(n+\frac{1}{2}))\xi^{2n+1}}} \right|_{-\sqrt{z_2}}^{+\sqrt{z_2}}\\
	\end{split}
	\ee
Then the off-shell superpotential is
	\be
	\label{1_off}
		\mathcal{W}^{\pm}=\pm\frac{c}{4\pi}{\sum_{n=0}^{\infty}\frac{(-1)^{n}\Gamma(1+3(n+\frac{1}{2}))^2(z_1)^{n+\frac{1}{2}}{}_2F_1(-1-2n,-\frac{1}{2}-n,\frac{1}{2}-n,z_2)}{(1+2n)\Gamma(1+n+\frac{1}{2})^4\Gamma(1+2(n+\frac{1}{2}))z_2^{n+\frac{1}{2}}}}
	\ee
And the on-shell superpotential at critical point $z_2=1$ is
	\be
	\label{1_on}
	\begin{split}
		W^{\pm}&=\left.\mathcal{W}^{\pm}\right|_{z_2=1}\\
		&=\pm\frac{c}{8}\sum_{n=0}^{\infty}\frac{\Gamma(1+3(n+\frac{1}{2}))^2(z_1)^{n+\frac{1}{2}}}{\Gamma(n+\frac{3}{2})^6}=\pm\frac{c}{8}B_{\{l\}}(z_1;\frac{1}{2})
	\end{split}
	\ee
\indent With above superpotential, we could further compute corresponding A-model instanton expansion after mirror map. The inverse mirror map in terms of $q_i=e^{t_i}$ is given by
	\be
		z_1=q_1-180 q_1^2+8910 q_1^3-948840 q_1^4-106787835 q_1^5+\dots.
	\ee
\indent Set the normalization constant $c=1$, then the instant expansion of superpotential encoding open Gromov-Written invariants is
	\be
		W(t_1)=18 \sqrt{q_1}+182 q_1^{3/2}+\frac{787968 }{25}q_1^{5/2}+\frac{323202744 }{49}q_1^{7/2}+\frac{15141625184 }{9}q_1^{9/2}+\mathcal{O}\left(q_1^{11/2}\right)
	\ee
And the Ooguri-Vafa invariants are given by:
	\be
		N_1 = 18,\ N_3 = 180,\ N_5 = 31518,\ N_7 = 6595974,\ N_9 = 1682402778,\ldots.
	\ee
These are agree with the J. Walcher's results\cite{Walcher:2009di}. Next we compute the open Ooguri-Vafa invariants of off-shell superpotential (\ref{1_off}). The off-shell superpotential instanton expansion has following form:
	\be
	\label{1_offins}
	\frac{\mathcal{W}(q)}{\omega_0(q)} = \sum_{k \atop {\rm odd}}\sum_{ d_1,d_2 \atop {\rm odd}}\frac{n_{d_1,d_2}}{k^2} q_1^{k d_1/2} q_2^{k d_2/2}
	\ee
The inverse open-closed mirror map are:
	\be
	\begin{split}
		&z_1 = q_1 -q_1 q_2 -180 q_1^2 -\frac{5}{4} q_1 q_2^2 +396 q_1^2 q_2 +8910 q_1^3 -\frac{29}{9} q_1 q_2^3 +312q_1^2 q_2^2+\ldots\\
		&z_2 = q_2 +2 q_2^2 +\frac{23}{4}q_2^3 -54q_1 q_2^2 +\frac{349}{18}q_2^4 -264q_1 q_2^3 +864 q_1^2 q_2^2+\ldots\\
	\end{split}
	\ee
Part of those invariants extracted from (\ref{1_offins}) are listed in TABLE I.
\begin{table}[!h]
\label{table1}
\def\temptablewidth{1.0\textwidth}
\begin{center}
\begin{tabular*}{\temptablewidth}{@{\extracolsep{\fill}}c|ccccc}
$d_1\backslash d_2$ &$-3$                    &$-1$                         &1        		   &3  	 			&5	    			  \\\hline
$-1$                                 &0			 &0			       & 0                        &0   	 			&0	    			  \\
1                                     &0            	  &$-9$    		     &$\frac{9}{2}$	     &$\frac{81}{8}$ 			&$\frac{431}{16}$   \\
3                                 &$\frac{1233}{8}$	 &$-\frac{14931}{16}$     &$\frac{15891}{64}$ &$\frac{29485}{384}$ &$-\frac{996505}{1024}$  \\
5                                 &$\frac{111366171}{1280}$ &$-\frac{1339110441}{5120}$ & $\frac{1933627533}{10240}$ &$\frac{4950353871}{81920}$ &$-\frac{23796622322211}{4096000}$ \\
7                                 &$\frac{14369812591077}{286720}$      &$-\frac{1797729647481}{16384}$ &$\frac{412319607217281}{4587520}$ &$\frac{22090306295992239}{32768000}$ &$-\frac{6049158010118187739}{917504000}$  \\
       \end{tabular*}
       {\rule{\temptablewidth}{1pt}}
\tabcolsep 0pt \caption{ Ooguri-Vafa invariants
$n_{(d_1,d_2)}$ for the off-shell superpotential $\mathcal{W}$
on the CICY in $\mathbb{P}^5[3,3]$. The horizonal
coordinates represent $d_2$ and vertical coordinates represent
$d_1$.} \vspace*{-12pt}
\end{center}
       \end{table}
\\
\\

\noindent{\it 2}. {\textbf{$\mathbb{P}^5_{211211}[4,4]$}}\\
\indent The vertices and nef-partition are:
	\be
		\begin{split}
			&E_1=\{ \nu^*_1=(1,0,0,0,0);\nu^*_2=(0,1,0,0,0);\nu^*_3=(0,0,1,0,0)\} \\
			&E_2=\{ \nu^*_4=(0,0,0,1,0);\nu^*_5=(0,1,0,0,1);\nu^*_6=(-2-1,-1,-2,-1) \}\\
		\end{split}
	 \ee
The charge vector is
	\be
		l=(-4,-4;2,1,1,2,1,1,),
	\ee	
and $z_1=\dfrac{a^2_{1,1}a_{1,2}a_{1,3}a^2_{2,1}a_{2,2}a_{2,3}}{a^4_{1,0}a^4_{2,0}}$. Similarly, we have
	\be
		\left \{
		\begin{split}
			&P_1=x_1^2+x_2^4+x_3^4+\psi x_4 x_5 x_6\\
			&P_2=x_4^2+x_5^4+x_6^4+\psi x_1 x_2 x_3\\
		\end{split}
		\right.
       	\ee
where $\psi=z_1^{-\frac{1}{8}}$.\\
\par
i) In the first case, we choose the divisor as $Q(D_1)=x_2^4+z_2x_3^4$, with critical value lying in $z_2=1$. In the corresponding enhanced polyhedron, where vertices are $E_1=\{\underline{\nu}^*_i=(1,0;{\nu}^*_i;0)\}_{i=1,2,3}$ and $E_2=\{\underline{\nu}^*_i=(0,1;{\nu}^*_i;0)\}_{i=4,5,6}$, this divisor is determined by following two vertices:
	\be
		\underline{w}_{1}^*=(1,0; 0,1,0,0,0;1),\ \underline{w}_{2}^*=(1,0; 0,0,1,0,0;1)
	\ee
This includes the curves chosen in the paper\cite{Walcher:2009di}. Then the subsystem is
	\be
   		\left \{
		\begin{split}
			&P_1^{D_1}=x_1^2+x_2^4+\widetilde{\psi} x_4 x_5 x_6\\
			&P_2^{D_1}=x_4^2+x_5^4+x_6^4+\widetilde{\psi}  x_1x_2^2\\
		\end{split}
		\right.
	\ee
where $\psi=u^{-\frac{1}{8}}$ and $u=(-\dfrac{z_1}{z_2})(1-z_2)^2$. Then the corresponding charge vector becomes
    \be
        \tilde{l}=(-4,-4;2,2,2,1,1).
    \ee
The GKZ operator of the Picard-Fuchs equation is:
	\begin{equation*}
		\mathcal{L}^D=(2\theta)^3 (2\theta-1)^3 \theta^2- u(4\theta-1)^2(4\theta-2)^2(4\theta-3)^2(4\theta-4)^2
	\end{equation*}
where $\theta=u\dfrac{\partial}{\partial u}$. Following the same steps, we could get
	\be
\pi(u;\frac{1}{2})=\frac{c}{2}\sum_{n=0}^{\infty}\frac{\Gamma(1+4(n+\frac{1}{2}))^2u^{n+\frac{1}{2}}}{\Gamma(1+2(n+\frac{1}{2}))^3\Gamma(1+n+\frac{1}{2})^2}=\frac{c}{2}B_{\{\widetilde l\}}(u;\frac{1}{2})
	\ee
Then the off-shell superpotential is
	\be
		\mathcal{W}^{\pm}=\pm\frac{c}{4\pi}\sum_{n=0}^{\infty}\frac{(-1)^{n}\Gamma(1+4(n+\frac{1}{2}))^2(z_1)^{n+\frac{1}{2}}{}_2F_1(-1-2n,-\frac{1}{2}-n,\frac{1}{2}-n,z_2)}{(1+2n)\Gamma(1+2(n+\frac{1}{2}))^3\Gamma(1+n+\frac{1}{2})^2z_2^{n+\frac{1}{2}}},
	\ee
and the on-shell superpotential
	\be
		W^\pm=\pm\frac{c}{8}\sum_{n=0}^{\infty}\frac{\Gamma(1+4(n+\frac{1}{2}))^2z_1^{n+\frac{1}{2}}}{\Gamma(1+2(n+\frac{1}{2}))^2\Gamma(1+n+\frac{1}{2})^4}=\pm\frac{c}{8}B_{\{ l\}}(z_1;\frac{1}{2}).
	\ee
The mirror map is given by
	\be
	\label{1.2_1mp}
		z_1=q_1-960q_1^2+213600q_1^3-160471040q_1^4-136981068240q_1^5+\ldots
	\ee
Then the instanton expansion of superpotential encoding open Gromov-Witten invariants is
	\be
	\begin{split}
		W=&64 \sqrt{q_1}+\frac{50176 q_1^{3/2}}{9}+\frac{116721664 q_1^{5/2}}{25}+\frac{275837288448 q_1^{7/2}}{49}+\mathcal{O}\left(q_1^{11/2}\right)\\
	\end{split}
	\ee
The extracted Ooguri-Vafa invariants are
	\be
		N_1 = 64,\ N_3 = 5568,\ N_5 = 4668864,\ N_7 = 5629332416,\ N_9 = 8291643121152,\ldots.
	\ee
These are still agree with J. Walcher's results\cite{Walcher:2009di}. We also give the open Ooguri-Vafa invariants of off-shell superpotential. The instanton expansion of  $\mathcal{W}$ in the A-model side:
	\be
		\frac{\mathcal{W}(q)}{\omega_0(q)} = \sum_{k \atop {\rm odd}}\sum_{ d_1,d_2 \atop {\rm odd}}\frac{n_{d_1,d_2}}{k^2} q_1^{k d_1/2} q_2^{k d_2/2}
	\ee
can be obtained through following mirror map:
	\be
	\begin{split}
		&z_1 = q_1 -q_1 q_2 -960q_1^2 -\frac{5}{4} q_1 q_2^2 +2064 q_1^2 q_2 +213600 q_1^3 -\frac{107}{36} q_1 q_2^3 +\ldots\\
		&z_2 = q_2 +2 q_2^2 +\frac{11}{2}q_2^3 -216q_1 q_2^2 +\frac{157}{9}q_2^4 -1008q_1 q_2^3 +3264 q_1^2 q_2^2+\ldots\\
	\end{split}
	\ee
Results are shown in Table II.
\begin{table}[!h]
\def\temptablewidth{1.0\textwidth}
\begin{center}
\begin{tabular*}{\temptablewidth}{@{\extracolsep{\fill}}c|ccccc}
$d_1\backslash d_2$ &$-3$                    &$-1$                         &1        		   &3  	 			&5	    			  \\\hline
$-1$                                 &0			 &0			       & 0                        &0   	 			&0	    			  \\
1                                     &0            	  &$-16$    		     &$8$	     &$16$ 			&$\frac{118}{3}$   \\
3                                 &$1424$	 &$-9216$     &$4544$ &$\frac{34072}{9}$ &$\frac{3968}{3}$  \\
5                                 &$\frac{23056384}{5}$ &$-\frac{76538368}{5}$ & $\frac{63725696}{5}$ &$\frac{200914624}{45}$ &$-\frac{259889106632}{1125}$ \\
7                                 &$\frac{77756465152}{5}$     &$-\frac{1309448679424}{35}$ &$\frac{239196554752}{7}$ &$\frac{1191285826811648}{7875}$ &$-\frac{12585274168672256}{7875}$  \\
       \end{tabular*}
       {\rule{\temptablewidth}{1pt}}
\tabcolsep 0pt \caption{Ooguri-Vafa invariants
$n_{(d_1,d_2)}$ for the off-shell superpotential $\mathcal{W}$
on the CICY in $\mathbb{P}^5_{211211}$. The horizonal
coordinates represent $d_2$ and vertical coordinates represent
$d_1$.} \vspace*{-12pt}
\end{center}
       \end{table}
\\
\\

ii) In this second case, we choose following two vertices in the enhanced polyhedron:
	\be
		\underline{w}_{1}^*=(1,0; 1,0,0,0,0;1),\ \underline{w}_{2}^*=(1,0; 0,1,0,0,0;1).
	\ee
They determine a different divisor: $Q(D_2)=x_2^4+z_2x_1^2$, which takes critical value at $z_2=1$ and contains three curves defined as:
	\be
		\mathcal{C}_\eta=\{x_1=\alpha_1x_2^2,\ x_4=\alpha_2x_5^2,\ x_6^5+\eta\psi^{\frac{5}{3}}(\alpha_1^4\alpha_2)^{\frac{1}{3}}x_2^4x_5=0\}
	\ee
where $\eta$ is a third root of $1$, {\it i.e.} $\eta^3=1$, and $\alpha_{1,2}^2=-1$. Obviously, there exists a $\mathbb{Z}_3$-Symmetry, in contrast to the $\mathbb{Z}_2$-Symmetry in the first case. This is quit different from proceeding examples and need a detailed discussion. The subsystem is given by:
	\be
   		\left \{
		\begin{split}
			&P_1^{D_2}=x_2^4+x_3^4+\psi^\prime x_4 x_5 x_6\\
			&P_2^{D_2}=x_4^2+x_5^4+x_6^4+\psi^\prime x_2^3x_3\\
		\end{split}
		\right.
	\ee
where $\psi^\prime=u^{-\frac{1}{8}}$ and $u=(-\dfrac{z_1}{z_2})(1-z_2)^3$. The corresponding charge vector becomes
	\be
		l^\prime=(-4,-4;3,1,2,1,1).
	\ee
The GKZ operator of the Picard-Fuchs equation is:
	\begin{equation*}
		\mathcal{L}^D=3\theta(3\theta-1)(3\theta-2) 2\theta(2\theta-1) \theta^3- u(4\theta-1)^2(4\theta-2)^2(4\theta-3)^2(4\theta-4)^2
	\end{equation*}
where $\theta=u\dfrac{\partial}{\partial u}$. This GKZ system have two solutions with fractional power:
	\be
	\begin{split}
		&\pi_1(u;\frac{1}{3})=c_1B_{\{l^\prime\}}(u;\frac{1}{3})=\sum_{n=0}^{\infty}\frac{\Gamma(1+4(n+\frac{1}{3}))^2u^{n+\frac{1}{3}}}{\Gamma(1+3(n+\frac{1}{3}))\Gamma(1+2(n+\frac{1}{3}))\Gamma(1+n+\frac{1}{3})^3}\\
		&\pi_2(u;\frac{2}{3})=c_2B_{\{l^\prime\}}(u;\frac{2}{3})=\sum_{n=0}^{\infty}\frac{\Gamma(1+4(n+\frac{2}{3}))^2u^{n+\frac{2}{3}}}{\Gamma(1+3(n+\frac{2}{3}))\Gamma(1+2(n+\frac{2}{3}))\Gamma(1+n+\frac{2}{3})^3}\\
	\end{split}
	\ee
They are dependent on $\xi:=z_2^{\frac{1}{3}}$. We consider $\pi_1(u;\frac{1}{3})$. In this case, the domain wall tension of curve $\mathcal{C}_\eta$ relative to a reference curve $\mathcal{C}_0$: $\mathcal{T}^{\eta,0}=\mathcal{W}^\eta-\mathcal{W}^0$ is
	\be
		\mathcal{T}^{\eta,0}=\frac{1}{2\pi i}\int_{\xi_0}^{\eta z_2^{\frac{1}{3}}}\pi_1(u,\frac{1}{3})\frac{3{\rm d}\xi}{\xi}
	\ee
where $\xi_0$ is a reference point. With this, domain wall tension between any two holomorphic curves could be computed. And I choose $\eta=\frac{1}{2}(-1+i\sqrt{3})$ during latter computation   In fact, we only need to compute the domain wall tension between two specific curves, while other cases could be obtained just by multiplying $\eta$ and $\eta^2$ because of the $\mathbb{Z}_3$-symmetry.
	\be
	\label{1.2_2T_offshell}
	\begin{split}
		&\frac{2\pi i}{c_1}\mathcal{T}^{\eta^2,\eta}=\frac{2\pi i}{c_1}(\mathcal{W}^{\eta^2}-\mathcal{W}^\eta)\\
		&=3\sum_{n=0}^{\infty}\frac{(-1)^{n+\frac{1}{3}}\Gamma(1+4(n+\frac{1}{3}))^2z_1^{n+\frac{1}{3}}{}_2F_1(-1-3n,-\frac{1}{3}-n,\frac{2}{3}-n,z_2)}{(1+3n)\Gamma(3n+2)\Gamma(2n+\frac{5}{3})\Gamma(n+\frac{4}{3})^3z_2^{n+\frac{1}{3}}}(\eta^2-\eta)
	\end{split}
	\ee
So the domain wall tension at critical point $z_2=1$: $T^{\eta^2,\eta}=\left.\mathcal{T}^{\eta^2,\eta}\right|_{z_2=1}$ is
	\be
	\label{1.2_2T_onshell}
		 T^{\eta^2,\eta}=\frac{c_1^{\prime}}{2\pi}(\eta^2-\eta)\sum_{n=0}^{\infty}\frac{\Gamma(1+4(n+\frac{1}{3}))^2z_1^{n+\frac{1}{3}}}{\Gamma(2n+\frac{5}{3})^2\Gamma(n+\frac{4}{3})^4}=\frac{c_1^{\prime}}{2\pi}(\eta^2-\eta)B_{\{l\}}(z_1;\frac{1}{3})
	\ee
In order to obtain the superpotential from this domain wall tension $\mathcal{T}^{\eta^2,\eta}=\mathcal{W}^{\eta^2}-\mathcal{W}^{\eta}$, we could always choose proper reference curve such that $\mathcal{W}^{\eta^2}=\eta^2\mathcal{W}^1$ and $\mathcal{W}^{\eta}=\eta\mathcal{W}^1$. From Eq.(\ref{1.2_2T_offshell}), the off-shell superpotentials could be read as:
	\be
	\begin{split}
		&\mathcal{W}^{1}(z_1,z_2;\frac{1}{3})=\frac{\tilde{c}_1}{2\pi}\sum_{n=0}^{\infty}\frac{(-1)^{n+\frac{1}{3}}\Gamma(1+4(n+\frac{1}{3}))^2z_1^{n+\frac{1}{3}}{}_2F_1(-1-3n,-\frac{1}{3}-n,\frac{2}{3}-n,z_2)}{(1+3n)\Gamma(3n+2)\Gamma(2n+\frac{5}{3})\Gamma(n+\frac{4}{3})^3z_2^{n+\frac{1}{3}}}\\
		&\mathcal{W}^{\eta}(z_1,z_2;\frac{1}{3})=\eta \mathcal{W}^1(z_1,z_2;\frac{1}{3}),\quad \mathcal{W}^{\eta^2}(z_1,z_2;\frac{1}{3})=\eta^2 \mathcal{W}^1(z_1,z_2;\frac{1}{3}) \\
	\end{split}
	\ee
And the on-shell superpotentials are
	\be
	\begin{split}
		&W^1(z_1;\frac{1}{3})=\frac{\tilde{c}_1^{\prime}}{2}B_{\{l\}}(z_1;\frac{1}{3})=\frac{\tilde{c}_1^{\prime}}{2}\sum_{n=0}^{\infty}\frac{\Gamma(1+4(n+\frac{1}{3}))^2z_1^{n+\frac{1}{3}}}{\Gamma(2n+\frac{5}{3})^2\Gamma(n+\frac{4}{3})^4}\\
		&W^{\eta}(z_1;\frac{1}{3})=\eta W^1(z_1;\frac{1}{3}),\quad W^{\eta^2}=\eta^2 W^1(z_1;\frac{1}{3})\\
	\end{split}
	\ee
\indent The $\pi_2(u;\frac{2}{3})$ case could be obtained in a similar way and we don't bother to do it here. It's easy to see that results in paper\cite{Walcher:2009di} are agree with ours up to a linear combination. Let $\tilde{c}_1^{\prime}=-2$, and the mirror map is also given by Eq.(\ref{1.2_1mp}). During the calculation of instanton expansion, we found only the combination of the from like
	\begin{equation*}
		W = \eta W^1(\frac{1}{3})+\eta^2 W^1(\frac{2}{3})
	\end{equation*}
has integral coefficients. One can easily verify that this on-shell superpotential have $\mathbb{Z}_3$-sysmmetry respect the curves we choose. The instant expansion is given by
	\be
W=108 \eta q_1^{\frac{1}{3}}+675\eta^2 q_1^{\frac{2}{3}}+\frac{23139}{4}\eta^4 q_1^{\frac{4}{3}}+\frac{1876608}{25} \eta^5 q_1^{\frac{5}{3}}+\frac{204393024}{49} \eta^7 q_1^{\frac{7}{3}}+\dots
	\ee
with Ooguri-Vafa invariants list:
	\be
		N_1=108,\ N_2 = 648,\ N_4 = 5616,\ N_5 = 75060,\ N_7 = 4171284,\ N_8 = 73727712,\ldots.
	\ee
On the other hand, the open Ooguri-Vafa invariants of the off-shell superpotential $\mathcal{W}=\eta \mathcal{W}^1+\eta^2 \mathcal{W}^2$ could be extracted from:
	\be
	\label{1.2_2_offshellexp}
		\frac{\mathcal{W}(q)}{\omega_0(q)} = \sum_{3\nmid k}\sum_{3\nmid d_1,d_2}\frac{n_{d_1,d_2}}{k^2} \eta^{k(d_1+d_2)} q_1^{k d_1/3} q_2^{k d_2/3}
	\ee
through the mirror map:	
	\be
	\begin{split}
		&z_1 = q_1 -2 q_1 q_2 -960q_1^2 -\frac{3}{2} q_1 q_2^2 +4272 q_1^2 q_2 +213600 q_1^3 -\frac{31}{9} q_1 q_2^3 +\ldots\\
		&z_2 = q_2 +2 q_2^2 -72 q_1 q_2+\frac{11}{2}q_2^3 -288q_1 q_2^2 -20820 q_1^2 q_2+\frac{157}{9}q_2^4+\ldots\\
	\end{split}
	\ee
TABLE III shows the open Ooguri-Vafa invariants extracted from Eq.(\ref{1.2_2_offshellexp}). Some of these numbers we get in this table are actually not integral and fractional anymore, we chose default significant figures in {\it Mathematica 10}, which is 6 digits, to show these numbers. Following the same steps as before, we believe this result is right.
\begin{table}[!h]
\def\temptablewidth{1.0\textwidth}
\begin{center}
\begin{tabular*}{\temptablewidth}{@{\extracolsep{\fill}}c|ccccc}
$d_1\backslash d_2$ &$-2$                    &$1$                         &4        		   &7  	 			&10	    			  \\\hline
1                                &0			 &0			       & 0                        &0   	 			&0	    			  \\
2                                &$-\frac{170497}{500}$  &$-\frac{12092}{25}$   &$\frac{153367}{100}$	     &$\frac{66282}{25}$ &$\frac{351157}{50}$   \\
4                                 &0	 &0     &0 &0 &0  \\
5                                 &$-\frac{345983}{5}$ &$\frac{167343}{2}$ & $-661883$ &$-443786$ &$4.70935\times 10^6$ \\
7                                 &0     &0 &0 &0 &0 \\
       \end{tabular*}
      \vspace*{12pt}
        {\rule{\temptablewidth}{1pt}}
\begin{tabular*}{\temptablewidth}{@{\extracolsep{\fill}}c|ccccc}
$d_1\backslash d_2$ &$-1$                   	        &$2$                         &5        		   &8 	 			&11	    			  \\\hline
1                                 &$-\frac{54414}{625}$ &$\frac{9069}{125}$& $\frac{12092}{125}$&$\frac{115949}{500}$&$\frac{342987}{500}$\\
2                                     &0            	  &0  		     &0	     &0 			&0   \\
4                                &$-\frac{210311}{10}$	 &$\frac{59733}{25}$     &$\frac{301371}{10}$ &$\frac{60816}{5}$ &$788999$  \\
5                                 &0 &0 & 0 &0 &0 \\
7                                 &$-2.74235\times 10^7$     &$3.60622\times 10^7$ &$1.05088\times 10^7$ &$-2.43337\times 10^8$ &$1.96372\times 10^9$  \\
       \end{tabular*}
       {\rule{\temptablewidth}{1pt}}
\tabcolsep 0pt \caption{Ooguri-Vafa invariants
$n_{(d_1,d_2)}$ for the off-shell superpotential $\mathcal{W}$
on the CICY in $\mathbb{P}^5_{211211}$. The horizonal
coordinates represent $d_2$ and vertical coordinates represent
$d_1$.} \vspace*{-12pt}
\end{center}
\end{table}
\\
\\
{\it 3}. {\textbf{$\mathbb{P}^5_{321321}[6,6]$}}\\
\indent In this system, we consider the nef-partition:
	\be
		\begin{split}
			&E_1=\{ \nu^*_1=(1,0,0,0,0);\nu^*_2=(0,1,0,0,0);\nu^*_3=(0,0,1,0,0)\} \\
			&E_2=\{ \nu^*_4=(0,0,0,1,0);\nu^*_5=(0,0,0,0,1);\nu^*_6=(-3,-2,-1,-3,-2) \}\\
		\end{split}
	 \ee
and the charge vector is
	\be
		l=(-6,-6;3,2,1,3,2,1).
	\ee
The corresponding defining polynomials in homogeneous coordiantes are:
	\be
		\left \{
		\begin{split}
			&P_1=x_1^2+x_2^3+x_3^6+\psi x_4 x_5 x_6\\
			&P_2=x_4^2+x_5^3+x_6^6+\psi x_1 x_2 x_3\\
		\end{split}
		\right.
       	\ee
with the $\psi= z_1^{-\frac{1}{12}}$ and $z_1=\dfrac{a_{1,1}^3 a_{1,2}^2 a_{1,3} a_{2,1}^3 a_{2,2}^2 a_{2,3} }{a_{1,0}^6 a_{2,0}^6}$. This case also have a situation with a $\mathbb{Z}_3$-symmetry by choosing the divisor as $Q(D)=z_2 x_2^3+x_3^6$. Then the calculation progress is identical to the former example. So we define the divisor by $Q(D)=z_2 x_1^2+x_3^6$ with critical value lying in $z_2=1$, which contains four curves with a $\mathbb{Z}_4$-symmetry:
	\be
		\mathcal{C}_{\zeta}=\{x_1=\alpha_1 x_3^3,\ x_4 = \alpha_2 x_6^3,\ x_2^2=\zeta \psi (\alpha_2^3 \alpha_1)^{\frac{1}{4}}x_3 x_6^3\},
	\ee
where $\alpha_1^2=\alpha_2^2=-1$ and $\zeta$ is one of the fourth-roots of $1$, {\it i.e.} $\zeta^4=1$. In addition, since it contains a nontrivial subgroup $\mathbb{Z}_2$, it is necessary to consider the representation of superpotential on $\mathbb{Z}_2$ induced from $\mathbb{Z}_4$.\\
\indent In subsystem, the $l$-vector reduced to
	\be
		l^\prime=(-6,-6;2,4,3,2,1).
	\ee
And the solutions with fractional powers could be obtained:
	\be
	\label{1.3_subper}
		\pi_1(u;\frac{1}{4})=c_1B_{\{l^\prime\}}(u;\frac{1}{4}),\quad \pi_2(u;\frac{1}{2})=c_2B_{\{l^\prime\}}(u;\frac{1}{2}),\quad \pi_3(u;\frac{3}{4})=c_3B_{\{l^\prime\}}(u;\frac{3}{4})
	\ee
Take $\pi_1(u;\frac{1}{4})$ for example. It is same as the above example of $\mathbb{Z}_3$-symmetry that we need to choose proper reference point corresponding to a reference curve $\mathcal{C}_0$, and we use $1,\ \zeta,\ \zeta^2$ and $\zeta^3$ to distinguish four $\mathbb{Z}_4$-symmetric curves. Define the new open moduli $\xi:=z_2^{\frac{1}{4}}$.
	\be
	\begin{split}
		\mathcal{T}^{\zeta,1}&=\mathcal{T}^{\zeta,0}-\mathcal{T}^{1,0}=\frac{1}{2\pi i}\int_{z_2^{\frac{1}{4}}}^{\zeta z_2^{\frac{1}{4}}}\pi_1(u(\xi);\frac{1}{4})\frac{4 {\rm d} \xi}{\xi}\\
		&=\frac{c_1^\prime}{2\pi}\sum_{n=0}^{\infty}\frac{\Gamma(6n+\frac{5}{2})^2(-1)^{n+\frac{1}{4}}z_1^{n+\frac{1}{4}}4 {}_2F_1(-1-4n,-\frac{1}{4}-n,\frac{3}{4}-n,z_2)}{\Gamma(4n+2)\Gamma(3n+\frac{7}{4})\Gamma(2n+\frac{3}{2})^2\Gamma(n+\frac{5}{4})(4n+1)z_2^{n+\frac{1}{4}}}\zeta^3(\zeta-1)
	\end{split}
	\ee
Choosing the proper reference curve such that $\mathcal{W}^{\zeta} = \mathcal{T}^{\zeta,0}=\zeta \mathcal{T}^{1,0}=\zeta \mathcal{W}^1$, so we could write the superpotential indexed by $1$:
	\be
		\mathcal{W}^{1}=\frac{c_1^\prime}{2\pi}\sum_{n=0}^{\infty}\frac{\Gamma(6n+\frac{5}{2})^2(-1)^{n+\frac{1}{4}}z_1^{n+\frac{1}{4}}4 {}_2F_1(-1-4n,-\frac{1}{4}-n,\frac{3}{4}-n,z_2)}{\Gamma(4n+2)\Gamma(3n+\frac{7}{4})\Gamma(2n+\frac{3}{2})^2\Gamma(n+\frac{5}{4})(4n+1)z_2^{n+\frac{1}{4}}}\zeta^3 \cdot 1,
	\ee
and similarly we could obtain those superpotential indexed by $\zeta$, $\zeta^2$ and $\zeta^3$ by:
	\be
	\label{1.3_z4}
		\mathcal{W}^{\zeta}=\zeta \mathcal{W}^1,\quad \mathcal{W}^{\zeta^2}=\zeta^2 \mathcal{W}^1,\quad \mathcal{W}^{\zeta^3}=\zeta^3 \mathcal{W}^1.
	\ee
\indent The on-shell superpotential is the off-shell superpotential at critical values $z_2=1$:
	\be
		W^1=\left.\mathcal{W}^1\right|_{z_2=1}=\frac{\tilde{c}_1}{2\pi}\sum_{n=0}^{\infty}\frac{\Gamma(6n+\frac{5}{2})^2 z_1^{n+\frac{1}{4}}}{\Gamma(3n+\frac{7}{4})^2\Gamma(2n+\frac{3}{2})^2\Gamma(n+\frac{5}{4})^2}=\frac{\tilde{c}_1}{2\pi}B_{\{l\}}(z_1;\frac{1}{4}),
	\ee
and this on-shell superpotential, together with $W^\zeta$, $W^{\zeta^2}$ and $W^{\zeta^3}$ from Eq(\ref{1.3_z4}), being a representation of $\mathbb{Z}_4$-symmetry. This procedure also hold for the other two solutions of Eq(\ref{1.3_subper}). The results are agree with J. Walcher's\cite{Walcher:2009di} up to a linear combination:
	\be
	\begin{split}
	&W_{\pm} = \pm W^2(z_1;\frac{1}{2}),\\
	&W_{\zeta}=\zeta W^1(z_1;\frac{1}{4})+\zeta^2 W^2(z_1;\frac{1}{2})+\zeta^3 W^3(z_1;\frac{3}{4}),\\
	\end{split}
	\ee
where the first superpotential is equipped with $\mathbb{Z}_2$-symmetry induced from $\mathbb{Z}_4$ and the second one with $\mathbb{Z}_4$-symmetry. Certainly, the instanton expansions coincident with the results in the paper\cite{Walcher:2009di}, and we don't list them here anymore. \\
\indent Next we give the open Ooguri-Vafa invariants of the off-shell superpotentials both $\mathcal{W}_{\pm}$ and $\mathcal{W}_{\zeta} = \zeta \mathcal{W}^1 +\zeta^2 \mathcal{W}^2 + \zeta^3 \mathcal{W}^3$. They share the same open-closed inverse map:
	\be
	\begin{split}
	&z_1 = q_1 - 3q_1 q_2-37440 q_1^2 -\frac{3}{4}q_1 q_2^2+244440q_1^2 q_2+\ldots \\
	&z_2 = q_2 -3000q_1 q_2 +2 q_2^2 -4800q_1 q_2^2 +\frac{11}{2}q_2^3+\ldots\\
	\end{split}
	\ee
And part of the open Ooguri-Vafa invariants for off-shell superpotential $\mathcal{W}_{\pm}$ and $\mathcal{W}_{\zeta}$ is listed in Table IV and Table V respectively.

	\begin{table}[!h]
\def\temptablewidth{1.0\textwidth}
\begin{center}
\begin{tabular*}{\temptablewidth}{@{\extracolsep{\fill}}c|ccccc}
$d_1\backslash d_2$ &-3    				&$-1$                   	 			&$1$                         				&3        		   					&5  	 		  \\\hline
-1				&0      				&0							&0  							&0								&0	\\
1                            	&0      				&-384			 			&192		      					&992   							& $\frac{5824}{3}$  			 \\
3                           	&611712				&-7602688           				&11227904    					&$\frac{263167552}{9}$	     			&$-\frac{131138752}{9} $  \\
5                            	&69062361088     		&-544235467008	 			&$\frac{4595731998848}{3}$     	&$-\frac{634121628352}{3}$    			&$-\frac{682615784470656}{125}$   \\
7                            	&9323856920051712     	&$-\frac{423421873354742784}{7}$ &$\frac{1459259189192285696}{7}$ & $-\frac{43329412223163433984}{175}$ &$-\frac{377055293849937948544}{525}$ \\
       \end{tabular*}
       {\rule{\temptablewidth}{1pt}}
\tabcolsep 0pt \caption{Ooguri-Vafa invariants
$n_{(d_1,d_2)}$ for the off-shell superpotential $\mathcal{W}_{\pm}$
on the CICY in $\mathbb{P}^5_{321321}$. The horizonal
coordinates represent $d_2$ and vertical coordinates represent
$d_1$.} \vspace*{-12pt}
\end{center}
       \end{table}

\begin{table}[!h]
\def\temptablewidth{1.0\textwidth}
\begin{center}
\begin{tabular*}{\temptablewidth}{@{\extracolsep{\fill}}c|ccccc}
$d_1\backslash d_2$ &$-3$                    &$1$                         &5      	  	 &9  	 			  \\\hline
1                                &0			 &0			       & 0                        &0   	 				  \\
2                                &0  &0   &0    &0   \\
3                                 &$-\frac{91641}{50}$	 &$-\frac{970151}{100}$     &$\frac{63129}{2}$ &$\frac{140011}{5}$   \\
5                                 &0 &0 &0 &0  \\
6                                 &0     &0 &0 &0  \\
7                                 &$-7.14106\times 10^7$     &$6.48917\times 10^7$&$6.05118\times 10^8$ &$-5.93448\times 10^8$  \\
       \end{tabular*}
      \vspace*{12pt}
        {\rule{\temptablewidth}{1pt}}

\begin{tabular*}{\temptablewidth}{@{\extracolsep{\fill}}c|ccccc}
$d_1\backslash d_2$ &$-2$                    &$2$                         &6        		   &10  	 			  \\\hline
1                                &0			 &0			       & 0                        &0   	 		  \\
2                                &$-\frac{350059}{1000}$  &$192$   &$\frac{960887}{1000}$	     &$\frac{5744}{3}$    \\
3                                 &0	 &0     &0 &0   \\
5                                 &0 &0 &0 &0  \\
6                                 &-7602688     &$1.12303\times 10^7$ &$\frac{263167552}{9}$ &$-1.45789\times 10^7$ \\
7                                &0     &0 &0 &0 \\
       \end{tabular*}
      \vspace*{12pt}
        {\rule{\temptablewidth}{1pt}}

\begin{tabular*}{\temptablewidth}{@{\extracolsep{\fill}}c|ccccc}
$d_1\backslash d_2$ &$-1$                   	 &$3$                         &7        		   &11 	 		 \\\hline
1                                 &$-\frac{27153}{200}$ &$\frac{124451}{1000}$& $\frac{78489}{500}$&$\frac{186971}{500}$\\
2                                     &0            	  &0  		     &0	     &0 		 \\
3                                &0	 &0     &0 &0   \\
5                                 &$-826742$ &$511632$ &$2.58187\times 10^6$ &$972622$  \\
6                                 &0     &0 &0 &0   \\
7                                &0     &0 &0 &0 \\
       \end{tabular*}
       {\rule{\temptablewidth}{1pt}}
\tabcolsep 0pt \caption{Ooguri-Vafa invariants
$n_{(d_1,d_2)}$ for the off-shell superpotential $\mathcal{W}_{\zeta}$
on the CICY in $\mathbb{P}^5_{211211}$. The horizonal
coordinates represent $d_2$ and vertical coordinates represent
$d_1$.} \vspace*{-12pt}
\end{center}
\end{table}

\subsection{Two-deformation Moduli Compact CICYs with D-branes}
\indent We consider two hypersurfaces, defined by degree $(4,0)$ and $(2,2)$ respectively, intersecting in the product space $\mathbb{P}_{21111}\times\mathbb{P}^1$:
	\begin{equation*}
		\left(\begin{array}{l}\mathbb{P}^1\\\mathbb{P}_{21111}\\\end{array}\right|\left|\begin{array}{cc}2&0\\2&4\\\end{array}\right)
	\end{equation*}
In the Toric geometry, the polytope corresponding to this system is given by following vertex:
	\begin{equation*}
	\begin{split}
		& \nu_1^*=(1,0,0,0,0),\ \nu_2^*=(0,1,0,0,0),\ \nu_3^*=(0,0,1,0,0),\ \nu_4^*=(0,0,0,1,0),\\
		& \nu_5^*=(-2,-1,-1,-1,0),\ \nu_6^*=(1,1,1,0,2),\ \nu_7^*=(-1,-1,-1,0,-2)\\
	\end{split}
	\end{equation*}
This Calabi-Yau can be resolved by adding a vertex $\nu_8^*=-\nu_1^*$. But the divisor $D_8$ defined by $\nu_8^*$ does not intersect with the complete intersection, so the component corresponding to $\nu_8^*$ can be dropped as we mentioned in last section. We recommend to refer more details in A. Klemm's paper\cite{Klemm:2004dl}. The nef-partition is chosen as $E_1=\{\nu_1^*,\nu_3^*,\nu_4^*\}$, $E_2=\{\nu_2^*,\nu_5^*,\nu_6^*,\nu_7^*\}$. This complete intersection Calabi-Yau manifold is defined by the following equations in toric coordinates:
		\be
		\left \{
		\begin{split}
			W_1(a,X)=&a_{1,0}+a_{1,1} X_1+a_{1,2} X_3+a_{1,3} X_4\\
			W_2(a,X)=&a_{2,0}+a_{2,1} X_2+a_{2,2} X_1^{-2} (X_2 X_3 X_4)^{-1}+a_{2,3} X_1 X_2 X_3 X_5^2+a_{2,4} (X_1 X_2 X_3)^{-1} X_5^{-2}\\
		\end{split}
		\right.
	\ee
And the relation lattice is generated by:
	\be
	\begin{tabular}{m{25pt}<{\centering}| m{25pt}<{\centering} m{20pt}<{\centering} m{20pt}<{\centering} m{20pt}<{\centering} m{20pt}<{\centering} m{20pt}<{\centering} m{20pt}<{\centering} m{20pt}<{\centering} m{20pt}<{\centering}}
	~           & $0_1$  &$0_2$  & $1 $  & $2$  & $3$  & $4$  & $5$  & $6$  & $7$ \\\hline
	${l}_1$  & $-4$    &$-2$     & $2$   & $1$  & $1$  & $1$  & $1$  & $0$  & $0$ \\
	${l}_2$  &$0$      &$-2$     & $0$   & $0$  & $0$  & $0$  & $0$  & $1$  & $1$ \\
	\end{tabular}
	\ee
Then two torus invariant closed-moduli parameters can be written out as:
	\be
	z_1=\frac{a_{1,1}^2 a_{2,1}a_{1,2}a_{1,3}a_{2,2}}{a_{1,0}^4 a_{2,0}^2},\quad z_2=\frac{a_{2,3} a_{2,4}}{a_{2,0}^2},
	\ee
 In the enhanced polyhedron, the vertices are $E_1=\{\underline{\nu}^*_i=(1,0;{\nu}^*_i;0)\}_{i=1,3,4}$ and $E_2=\{\underline{\nu}^*_i=(0,1;{\nu}^*_i;0)\}_{i=2,5,6,7}$, and we choose two vertices in the enhanced polyhedron:
	\be
		\underline{w}_{3}^*=(1,0; 0,0,1,0,0;1),\ \underline{w}_{4}^*=(1,0; 0,0,0,1,0;1),
	\ee
which determine the divisor: $Q(D)=b_1 X_3+b_2 X_4$. The open modulus $z_3=\dfrac{b_2}{b_1}$, since $X_3$ and $X_4$ correspond to homogeneous coordinates with same weight. The subsystem is described by
\be
\begin{tabular}{m{25pt}<{\centering}| m{25pt}<{\centering} m{20pt}<{\centering} m{20pt}<{\centering} m{20pt}<{\centering} m{20pt}<{\centering} m{20pt}<{\centering} m{20pt}<{\centering} m{20pt}<{\centering}}
	   ~                 & $0_1$ &$0_2$   &$1 $ &$2$   & $3$  & $4$  & $5$  & $6$  \\\hline
	$\tilde{l}_1$   & $-4$    &$-2$     &$2$   &$2$   & $1$  & $1$  & $0$  & $0$ \\
	$\tilde{l}_2$   &$0$      & $-2$    & $0$  &$0$   & $0$  & $0$  & $1$  & $1$ \\
\end{tabular}
\ee
The GKZ operator of the Picard-Fuchs equation is:
	\begin{equation*}
	\begin{split}
		&\mathcal{L}_1^D=\\
		&4\theta_1^4(2\theta_1-1)^2- u_1(4\theta_1-1)(4\theta_1-2)(4\theta_1-3)(4\theta_1-4)(2\theta_1+2\theta_2-1)(2\theta+2\theta_2-2),\\
		&\mathcal{L}_2^D=\theta_2^2-u_2 (2\theta_1+2\theta_2-1)(2\theta+2\theta_2-2)\\
	\end{split}
	\end{equation*}
where $\theta_1=u_1\dfrac{\partial}{\partial u_1},\ \theta_2=u_2\dfrac{\partial}{\partial u_2}$. Similarly to the first example, we could obtain the special solution for this subsystem:
	\be
		\pi_{\{\tilde{l}\}}(u_1,u_2;\frac{1}{2},0) =c \sum_{n=0}^{\infty}\frac{\Gamma(1+4(n_1+\frac{1}{2}))\Gamma(1+2(n_1+\frac{1}{2})+2n_2)u_1^{n_1+\frac{1}{2}}u_2^{n_2}}{\Gamma(1+2(n_1+\frac{1}{2}))^2\Gamma(1+(n_1+\frac{1}{2}))^2\Gamma(1+n_2)^2}
	\ee
The off-shell superpotential by integrating it is:
	\be
	\begin{split}
		&\mathcal{W}=\\
		&\tilde{c} \sum_{n=0}^{\infty}\frac{(-1)^{n_1+\frac{1}{2}}\Gamma(4n_1+3)\Gamma(2n_1+2n_2+2)z_1^{n_1+\frac{1}{2}}z_2^{n_2} {}_2F_1(-1-2n_1,-\frac{1}{2}-n_1,\frac{1}{2}-n_1,z_3)}{\Gamma(2n_1+2)^2\Gamma(n_1+\frac{3}{2})^2\Gamma(1+n_2)^2(1+2n_1) z_3^{n_1+\frac{1}{2}}}
	\end{split}
	\ee
And the on-shell superpotential is:
	\be
	W=\mathcal{W}\vert_{z_3 = 1}={c^{\prime}} \sum_{n=0}^{\infty}\frac{\Gamma(4n_1+3)\Gamma(2n_1+2n_2+2)z_1^{n_1+\frac{1}{2}}z_2^{n_2}}{\Gamma(2n_1+2)\Gamma(n_1+\frac{3}{2}))^4\Gamma(1+n_2)^2}
	\ee
First, for on-shell superpotential calculation, the inverse mirror map is given by:
	\be
	\begin{split}
		z_1=&q_1-6q_1 q_2 -104 q_1^2  +17 q_1 q_2^2+448q_1^2 q_2+ 6444 q_1^3 -32 q_1 q_2^3-816 q_1^2 q_2^2 +\ldots\\
		z_2=&q_2-2q_2^2-72 q_1 q_2 +3q_2^3 +288 q_1 q_2^2+1308q_1^2 q_2 -4 q_2^4 -576 q_1 q_2^3 +\ldots\\
	\end{split}
	\ee
The Ooguri-Vafa invariants of on-shell superpotential are listed below.
\begin{table}[!h]
\def\temptablewidth{1.0\textwidth}
\begin{center}
\begin{tabular*}{\temptablewidth}{@{\extracolsep{\fill}}c|ccccc}
$d_1\backslash d_2$ &0                    &1                         &2         		   &3   	 			&4	    			  \\\hline
1                                 & 64                 &64         		     & 0      			   &0   	 			&0 	    			  \\
3                                 & -320              &5888     		     &5888 			   &320    				&0       			  \\
5                                 &2880 		 &625792 		     & 4040192		   &4040192 			&625792 			  \\
7                                 &-35392 		 &63595008 	     & 1247366656 	   &4318406144 		&4318406144 		  \\
9                                 &506432          &6250007232       &276467300352 	   &2233590214912 		&5775335092224 	  \\
11		                  &15768871104 &1069086628608 &94908374836480 &1326149570899200  	&6214165527984128 \\
       \end{tabular*}
       {\rule{\temptablewidth}{1pt}}
\tabcolsep 0pt \caption{Ooguri-Vafa invariants
$N_{(d_1,d_2)}$ for the on-shell superpotential $W$
on the CICY in $\mathbb{P}_{21111}\times\mathbb{P}^1$. The horizonal
coordinates represent $d_2$ and vertical coordinates represent
$d_1$.} \vspace*{-12pt}
\end{center}
\end{table}

For the off-shell superpotential, the inverse mirror map is given by:
	\be
	\begin{split}
		z_1=&q_1-104q_1^2 -6 q_1 q_2 -q_1q_3+6444q_1^3+488q_1^2q_2+232q_1^2q_3+17q_1q_2^2-\frac{3}{4}q_1q_3^2+\ldots\\
		z_2=&q_2-2q_2^2-72q_1 q_2+3q_2^3+1308q_1^2q_2+288q_1q_2^2+72q_1q_2q_3-4344q_1^2q_2q_3 +\ldots\\
		z_3=&q_3-q_3^3+2q_1 q_3^2+12q_1q_3^3+144q_1^2 q_3^2+48q_1 q_2 q_3^2 -2q_2 q_3^3-546q_1^2 q_3^3+\ldots\\
	\end{split}
	\ee
\indent Part of the open Ooguri-Vafa invariants of this off-shell superpotential are listed in Table VII.

\begin{table}[!h]
\def\temptablewidth{1.0\textwidth}
\begin{center}
\begin{tabular*}{\temptablewidth}{@{\extracolsep{\fill}}c|cccccc}
$d1=1 \ d_2\backslash d_3$	&-1                    	&1		   	  	&3      	  	 	&5  	 			&7		  \\\hline
0                                		&-32				&-16                     	&-8   	 		&$\frac{32}{9}$		&$-\frac{125}{3}$	  \\
1                                		&-32   			&-16				&-40   			&$\frac{176}{9}$	&$-\frac{485}{3}$	  \\
2                                 		&0     			&0				&-64				&32				&-288   \\
3                                		&0     			&0				&-64				&32				&-384   \\
4                                 		&256 			&128				&32				&$-\frac{6160}{9}$	&$-\frac{3104}{3}$\\
5                                 		&2432			&1216 			&864 			&$-\frac{50240}{9}$  &$-\frac{11668}{3}$\\
\end{tabular*}
\vspace*{12pt}
{\rule{\temptablewidth}{1pt}}

\begin{tabular*}{\temptablewidth}{@{\extracolsep{\fill}}c|cccccc}
$d1=3 \ d_2\backslash d_3$	&-1                    	&1		   	  		&3      	  	 		&5  	 				&7		  \\\hline
0                                		&$-\frac{1664}{3}$	&$\frac{2176}{3}$  		&$\frac{2192}{27}$	         &$\frac{27848}{27}$	    	&$\frac{3992}{27}$	  \\
1                                		&-9344   			&$\frac{30976}{3}$		&-608			 	&7512				&$\frac{49984}{9}$	  \\
2                                 		&-9344   			&$\frac{54016}{3}$		&-1504				&$\frac{103048}{3}$		&$\frac{227104}{9}$	  \\
3                                		&$-\frac{1664}{3}$	&$\frac{50816}{3}$  		&$-\frac{40432}{27}$		&$\frac{2089928}{27}$	&$\frac{1630712}{27}$	  \\
4                                 		&146816			&$-\frac{294848}{3}$		&$\frac{404416}{3}$		&12008				&$\frac{941192}{3}$\\
5                                 		&1707136			&$-\frac{4130624}{3}$	&$\frac{4917568}{3}$	&$-\frac{4264376}{3}$  	&$\frac{39350056}{9}$\\
\end{tabular*}
\vspace*{12pt}
{\rule{\temptablewidth}{1pt}}

\begin{tabular*}{\temptablewidth}{@{\extracolsep{\fill}}c|cccccc}
$d1=5 \ d_2\backslash d_3$	&-1                    			&1		   	  			&3      	  	 			&5  	 					&7		  \\\hline
0                                		&-53952					&$\frac{322336}{5}$			&$-\frac{562128}{5}$			&$\frac{11971712}{135}$		&$\frac{15841322}{27}$\\
1                                		&$-\frac{14700352}{5}$		&$\frac{22330208}{5}$		&$-\frac{241186096}{45}$   	&$\frac{501011024}{135}$		&$-\frac{544454426}{135}$	  \\
2                                 		&$-\frac{70166272}{5}$		&$\frac{111636608}{5}$		&$-\frac{433060288}{15}$		&$\frac{2809761344}{135}$	&$-\frac{659146360}{27}$	  \\
3                                		&$-\frac{87369472}{5}$		&$\frac{160379008}{5}$		&$-\frac{878087104}{15}$		&$\frac{7024433984}{135}$	&$-\frac{11830811096}{135}$	  \\
4                                 		&$\frac{34126592}{3}$		&$-\frac{360217216}{15}$		&$-\frac{1629777856}{135}$	&$\frac{1153591328}{135}$	&$-\frac{5773698856}{27}$\\
5                                 		&$\frac{2138051584}{3}$		&$-\frac{19871694848}{15}$	&$\frac{220223851072}{135}$	&$-\frac{73821754288}{45}$  	&$\frac{21373604864}{15}$\\
\end{tabular*}
{\rule{\temptablewidth}{1pt}}
\tabcolsep 0pt \caption{Ooguri-Vafa invariants
$n_{(d_1,d_2)}$ for the off-shell superpotential $\mathcal{W}$
on the CICY in $\mathbb{P}_{21111}\times\mathbb{P}^1$. The horizonal
coordinates represent $d_2$ and vertical coordinates represent
$d_1$.} \vspace*{-12pt}
\end{center}
\end{table}

\section{Summary}
The D-brane superpotential computation is essential for both physics and mathematics. Physically speaking, in the low-energy effective theory, superpotential determine the vacuum structure. In the view of A-model, it is the generating function of the Ooguri-Vafa Invariants of the Calabi-Yau manifold involving D-branes. And these Ooguri-Vafa Invariants are closely related to the number of the BPS states. Mathematically speaking, they count the holomorphic disks on Calabi-Yau manifolds. For the A-model on non-compact Calabi-Yau manifolds wrapped by D-branes, there are some straight methods, like localization, to calculate D-brane superpotential; while for the A-model on compact ones, it's very hard to calculate the superpotential directly, because they are non-perturbative. However, mirror symmetry and algebraic geometry method offer an effective approach to resolve this problem.\\
\indent In this paper, we further generalize the GKZ method to calculate on-shell and off-shell D-brane superpotentials on more general complete intersection Calabi-Yau manifolds with applications on several one-parameter and two-parameter models. And we also calculate both the on-shell and off-shell superpotential on the B-model side, and through open-closed mirror symmetry, we give non-perturbative superpotentials in the A-model side represented by the Ooguri-Vafa invariants. The on-shell results of one-deformation modulus compact CICYs with D-branes are in exactly agreement with J. Walcher's and M. Aganagic's results \cite{Walcher:2009di,Aganagic:2009jq} which are from different approaches.\\
\indent It is important for String/M/F-theory phenomenology and geometric invariants to study the D-brane superpotential for compact Complete Intersection Calabi-Yau manifolds (CICY)  with many open and closed string deformation moduli parameters and with the parallel and/or intersectiong D-branes wrapying on the holomorphic submanifolds of the compact CICY, and to study mathematical significance, such as enhanced gauge symmetries and geometric singularities\cite{Aspinwall:1995gn,Katz:1996ht,Sen:1996cf,Bershadsky:1996iq,Perevalov:1997ea,Candelas:1997fm}, etc. Furthermore, it is interesting to calculate the D-brane superpotential from the $\mathcal{A}_{\infty}$-structure in the derived category of coherent sheaves of Calabi-Yau manifold\cite{Aspinwall:2004wx,Aspinwall:2006vv,Aspinwall:2006ud,Aspinwall:2010wm}.\\

\section{Acknowledgement}
\indent This work is supported by NSFC(11475178) and Y4JT01VJ01.

\end{document}